# *IRIS* observational approach to the oscillatory and damping nature of network and internetwork chromosphere small-scale brightening (SSBs) and their unusual dynamical and morphological differences in different regions on the solar disk

**Rayhane Sadeghi[1]** · **Ehsan Tavabi*[1]**

**Abstract** One of the most exciting benefits of solar small-scale brightening is their oscillations, this study investigated the properties of small-scale brightening (SSBs) in different regions of the Sun and found that there are differences and similarities in the properties of oscillated and non-oscillated SSBs in different regions of the Sun, including quiet Sun (QS), the adjacent to active regions (AAR), and coronal hole (CH).

The damping per period ($Q$-factor) and maximum Doppler velocity of SSBs var- ied depending on the region, with the less bright internetwork SSBs in QS having lower damping time (120 seconds) and higher maximum Doppler velocities (47 km/s) compared to the brighter network SSBs (with 216 seconds & 37 km/s, respectively), while in AAR, internetwork SSBs tend to have higher damping time (about of 220 seconds) and wider maximum Doppler velocity (10 to 140 km/s) ranges compared to network SSBs (130 seconds & 10 to 85 km/s). In CH, both types of SSBs show similar damping time (120 seconds), but internetwork SSBs tend to have higher maximum Doppler velocities (100 km/s) compared to network SSBs (85 km/s).

Also, it was pointed out that the majority of network SSBs in AARs are in the overdamping mode, while in QS, internetwork SSBs demonstrate overdamping behavior and oscillated network SSBs exhibit critical damping behavior. It is important to bear in mind, however, that the physical mechanisms underlying the damping of SSBs may vary depending on the local plasma conditions and magnetic environment.

R. Sadeghi
rayhane.sadeghi@gmail.com
E. Tavabi*
e_tavabi@pnu.ac.ir

[1]  Physics Department, Payame Noor University, Tehran, Iran, 19395-3697

**Keywords:** Chromosphere network, Dynamics; Damping, Magnetic; Magnetic fields, Corona

## 1. Introduction

Bright Points (BPs) are rather small-scale magnetic structures identified in the solar photosphere and chromosphere that play an important role in energy transmission and solar atmospheric heating (Muller, 1973; Falconer et al., 1998; Berger et al., 1996). Their oscillatory behavior has been a topic of interest for several decades, with the first observations of oscillations in BPs reported by (Nesis et al., 2001). Despite extensive studies conducted on this subject, the relationship between oscillations and the properties of BPs remains a topic of ongoing research.

Kayshap et al. (2018b) used IRIS observations to study the propagation of photospheric waves into the chromosphere. They found that acoustic waves with periods of 1.6 to 4.0 minutes, successfully propagate into the chromosphere, with some locations showing the propagation of 5-minute oscillations. Kayshap et al. (2019) further investigated wave propagation within an active-region plage and found that slow magneto-acoustic waves (SMAWs) with periods between 2.0 and 9.0 minutes are correlated between the photosphere and transition region. These studies collectively suggest that a range of waves, including acoustic and SMAWs, play a significant role in the propagation of photospheric waves into the chromosphere.

BPs exhibit oscillatory behavior that is often associated with the emergence of new magnetic flux and the cancellation of existing flux (Ugarte-Urra et al., 2004; Tavabi et al., 2015). These oscillations can be observed in different period ranges. The oscillations are thought to be caused by propagating magneto-acoustic waves in loop systems associated with the BPs, or by recurrent magnetic reconnection (Tian et al., 2008; Ajabshirizadeh, Tavabi, and Koutchmy, 2008). Additionally, Alfv´en waves, produced by a torsional twist, have been detected in the lower solar atmosphere above BPs, with energy flux sufficient to heat the solar corona (Jess et al., 2009b; Tavabi et al., 2014). Coronal BPs exhibit quasi-periodic oscillations, which are linked to magnetic flux changes (Samanta, Pant, and Banerjee, 2015)). These oscillations are caused by propagating slow magneto-acoustic waves and standing slow waves in the solar transition region (Ajabshirizadeh, Tavabi, and Koutchmy, 2008; Sangal et al., 2022b). The oscillatory behavior is particularly prominent above BP-like structures in the quiet Sun (Zeighami, Tavabi, and Amirkhanlou, 2020a,b; Andic et al., 2010). The presence of subarcsecond BPs and quasi-periodic upflows in a quiescent filament channel further supports the role of small-scale oscillatory magnetic reconnections (Li and Zhang, 2016). Interface Region Imaging Spectrograph (*IRIS* ) BPs are tiny bright features that may be detected at the interface region between the Sun's photosphere and the corona. These BPs were detected in data conducted by NASA's *IRIS* in 2013 (Hou et al., 2016; Tavabi, 2018; De Pontieu et al., 2021; Kayshap and Dwivedi, 2017; Madjarska, 2019).

The interaction between the magnetic field and plasma can lead to plasma heating through various mechanisms. Pustovitov (2011) highlights the role of this interaction in the rapid redistribution of energy in magnetically confined equilibrium plasma. Shukla, Shukla, and Stenflo (2009) further explores this, demonstrating the generation of magnetic fields in warm plasma by the non-stationary ponderomotive force of an electromagnetic wave. Vodopyanov et al. (2020) extends this understanding to the simulation of plasma flow interaction with arched magnetic fields, a process relevant to solar phenomena and the Earth's magnetosphere. Lastly, Willett and Maraghechi (1977) examines the excitation of electron plasma waves by the interaction of electromagnetic waves in a magnetized plasma, providing a formula for the power absorbed per unit volume of plasma and studying the effects of the magnetic field on the plasma heating rate. The BPs observed by *IRIS* are the result of magnetic reconnection and rearrangements in the solar atmosphere, leading to the heating of the plasma (Zhao et al., 2017; Reale et al., 2019). These interactions can occur between pre-existing and emerging magnetic fields, as well as within the emerging flux region itself (Guglielmino et al., 2018). The heating process can involve both thermal and nonthermal mechanisms, with the latter potentially playing a significant role in the impulsive heating of the plasma (Reale et al., 2019). The BPs exhibit different characteristics in different temperature ranges, suggesting a two-stage heating process (Tian et al., 2008). These BPs may also play a role in the creation of other structures in the Sun's atmosphere, such as coronal loops (Sadeghi and Tavabi, 2022b; Tian et al., 2008; Zeighami, Tavabi, and Amirkhanlou, 2020a). Scientists expect to obtain a better knowledge of how the Sun's magnetic field and plasma interact, as well as how energy is distributed throughout the Sun's atmosphere, by examining *IRIS* BPs. *IRIS* data has already yielded fresh insights into the dynamics of the Sun's interface region, as well as the function of the magnetic field in creating the observed features (Parks, 2019; Zeighami, Tavabi, and Amirkhanlou, 2020a; Tavabi, 2018; Tavabi, Zeighami, and Heydari, 2022; Guglielmino et al., 2018; Tavabi and Sadeghi, 2024; Kayshap et al., 2020). *IRIS* BPs are important for practical reasons in addition to their scientific functionality. These features' energy can have an impact on the Earth's atmosphere and damage communication and navigation systems. Understanding how these features function will assist scientists in better predicting space weather and mitigating its impacts on our technologies (Jansen and Pirjola, 2004; Pirjola et al., 2005; Chandrasekhar et al., 2012) .

These characteristics have been seen to show oscillations or periodic fluctuations in brightness over time. These oscillations can tell us a lot about how magnetic fields and plasma behave in the Sun's atmosphere (Sadeghi and Tavabi, 2022a; Tavabi and Sadeghi, 2024; Sadeghi and Tavabi, 2024). Research on solar BPs has revealed a significant percentage of oscillations. Gao et al. (2022) found that 16 out of 23 coronal BPs exhibited decayless kink oscillations, with periods ranging from 1 to 8 minutes. Similarly, Ugarte-Urra et al. (2004) observed oscillatory behavior in the transition region lines of two BPs, with periods of 420-650 seconds and 491 seconds. Jess et al. (2009b) detected Alfv´en waves in a large bright-point group, with periodicities of 126-700 seconds. Tian et al. (2008) also identified long-period oscillations in solar coronal BPs, with periods ranging from 8 to 64 minutes. These studies collectively suggest that oscillations are a common feature of solar BPs. Research on the period of oscillation in solar BPs has revealed a range of findings. Tian et al. (2008) identified oscillations with periods ranging from 8 to 64 minutes in coronal BPs, with the cause still uncertain. Gao et al. (2022) reported decayless kink oscillations in these BPs, with periods ranging from 1 to 8 minutes. Ugarte-Urra et al. (2004) found a linear relationship between the appearance of coronal emission and the emergence of new magnetic flux, with some BPs exhibiting damped oscillations. Samanta, Pant, and Banerjee (2015) observed quasi-periodic brightenings in coronal BPs, linking them to underlying magnetic flux changes and transition region explosive events. These studies collectively suggest a complex interplay of factors influencing the period of oscillation in solar BPs.

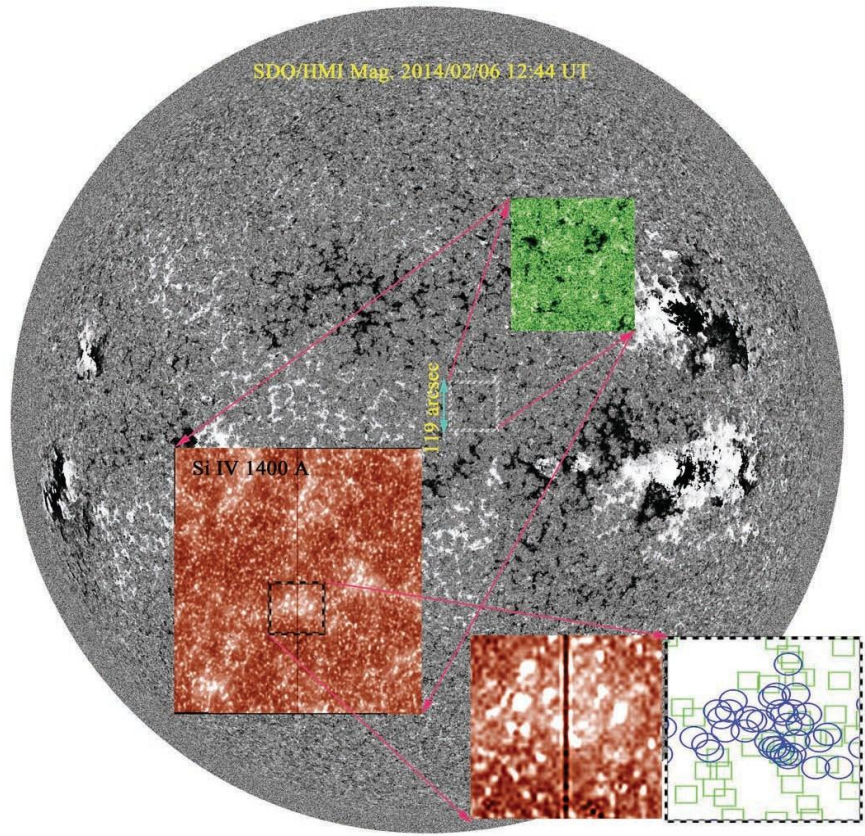

**Figure 1.** High-resolution image of the Sun's disk captured by the Helioseismic and Magnetic Imager (HMI) on February 6, 2014, at 12:44 UT. The location of the solar disk is magnified. The aligned 1400 Åwavelength image, related to Si IV, is overlaid onto the magnified region, highlighting the presence of SSBs. Using deep learning algorithms, a specific area of the 1400 Åimage is magnified and fed into the code, resulting in the identification and illustration of SSBs on the solar surface. The blue circles remark the network SSBs and the green rectangles are internetwork SSBs.

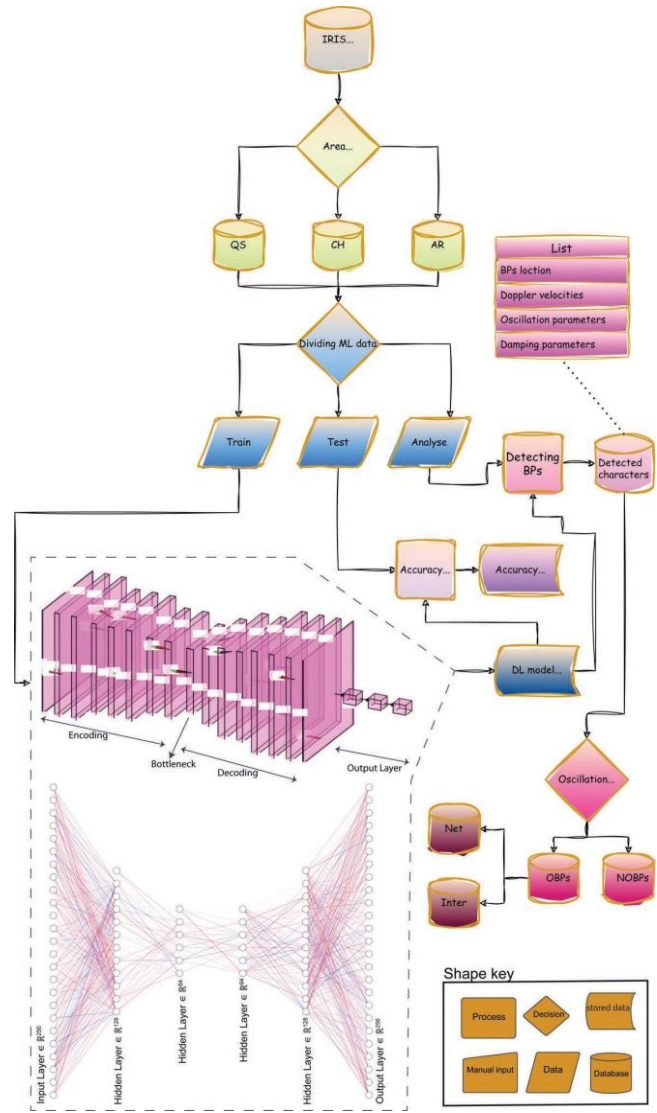

**Figure 2.** The chart describes the development of an image processing pipeline to identify and track SSBs in solar observations. The pipeline includes several steps, such as image enhancement, segmentation, feature extraction, and tracking. Machine learning algorithms are used to accurately identify and characterize the SSBs. The SSBs are classified into oscillated and non-oscillated groups based on the presence or absence of significant oscillatory power in their intensity time series. Wavelet analysis and Fourier analysis are employed to investigate the frequency and time-dependent properties of the oscillations. The statistical properties of network and internetwork SSBs are analyzed and compared. The study also discusses the development of a machine learning-based method for detecting bright features in solar images, which involves several steps including collecting annotated training data, designing and training a convolutional neural network, and evaluating its performance using metrics such as precision, recall, and F1-score.

Small amplitude oscillations in solar BPs, such as those observed in prominences, are subject to various damping mechanisms. These include thermal effects, mass flows, resonant damping in non-uniform media, and partial ionization effects (Arregui et al., 2010). In the case of solar-type stars, acoustic mode damping and excitation are influenced by stochastic processes (Houdek, 2006). The presence of oscillations in coronal BPs has been observed, with some displaying a damped oscillatory behavior (Ugarte-Urra et al., 2004). These findings suggest that a combination of physical mechanisms, including those related to the local plasma environment and the broader stellar context, contribute to the damping of oscillations in solar BPs.

Another issue that must be considered is the similarity between *IRIS* BPs and solar ultraviolet bursts. The term "solar ultraviolet bursts" refers to small, intense, transient brightenings in ultraviolet images of solar active regions (Young et al., 2018). These bursts are associated with small-scale, canceling opposite-polarity fields in the photosphere, emerging flux regions, and sunspot moats (Young et al., 2018; Peter et al., 2014). They are also linked to magnetic null points or bald patches (Chen et al., 2019; Chitta et al., 2017). The bursts exhibit enhanced and broadened spectral profiles in transition region lines, often with chromospheric absorption lines (Chen et al., 2019). Guglielmino et al. (2018) further characterized the UV emission properties of these bursts, highlighting their association with plasma ejections and small-scale eruptive phenomena. The properties of these bursts are similar to those of subarcsecond BPs and quasi-periodic upflows observed by the *IRIS* (Li and Zhang, 2016). The BPs are associated with magnetic reconnection events in the low solar atmosphere (Li and Zhang, 2016). The BPs in emerging flux regions also exhibit responses in lines formed from the upper photosphere to the transition region, with different heating mechanisms (Zhao et al., 2017). Therefore, the solar ultraviolet bursts discovered by *IRIS* are likely to be a type of BP, as they share similar properties and are associated with magnetic reconnection events.

By evaluating the structure of magnetic fields and the presence of these oscillations at solar BPs, scientists can gain important insights into the mechanisms that regulate the Sun's activity. They may also use the observational data to validate and refine models of the Sun's magnetic field and plasma (Priest, 1983; Weiss and Tobias, 2000; Erd´elyi, 2006).

The network BPs, as the name implies, are located mostly in the Sun's network regions, which have high magnetic fields. They form a network of brilliant spots that outline the magnetic field arrangement. Inter-network BPs, on the other hand, arise in zones of the Sun with lower magnetic fields - thus the name "inter-network". These weak flash-like points had been neglected for decades, but with new and better high-resolution telescopes, they may finally be studied in detail (Muller et al., 1994; Almeida et al., 2004).

While solar BPs are known to oscillate, this is not true for all BPs. In reality, different types of solar BPs have diverse features, and while some oscillate, others do not (Ugarte-Urra et al., 2004; Gao et al., 2022; Khomenko and Collados, 2015).

CHs are one of the darkest regions in the solar atmosphere even in comparison to the QS at the coronal temperatures. In addition, at TR temperatures, QS and CHs are indistinguishable. Kayshap et al. (2019) used the Mg II 2796.35 Åspectral line taken by the IRIS to reveal the similarities and differences in the QS and CH at TR and low corona layers. Their investigation illustrated that the Mg II k3 & k2v emission lines that originate mainly in the chromosphere are significantly lower in CH than in QS for the regions with similar magnetic field strength. The wing emissions of Mg II k that originate from the photospheric layer, however, do not show any difference between QS and CH. In addition, they reported the variations in Mg II k3 intensities between QS and CH increase with growing magnetic field strength. Thus, typical variations in magnetic flux (or a distribution of the flux) in the CH relative to the QS could explain any differences anchored in the lower atmosphere. The intensities in the k2 and k3 of the Mg II line are directly related to the magnetic field (Leenaarts et al. 2013a). They found that the bright patches in the TR are located above the chromospheric network cells where the magnetic field is rather stronger. The darker points in the TR are above the inter-network regions where the magnetic field is weaker. the network region (bright patch) is surrounded by a strong magnetic field, propagating waves in this region may carry more energy than propagating waves in the inter-network region (dark patch), which is surrounded by a weak magnetic field. Those results are not only important for mass and energy supply from the chromosphere to the corona but also provide essential ingredients for the modeling of the solar spectral irradiance for the understanding of the space-climate relationships. The radiation change between QS and CH is one the most significant concepts to explain temperature variation through these layers and was explained by Wiegelmann & Solanki (2004) by invoking loop statistics and Rosner-Tucker-Vaiana (RTV; Rosner et al. (1978) scaling, and more recently using the space-born data by Kayshap et al. (2019). In this research, fully aware of these differences, we decided to concentrate on the periods of oscillation of Doppler velocity and intensity in these three regions (CH, QS, and adjacent active regions), and their damping ratios.

In this research, the term "small-scale brightenings" (SSBs) was utilized in the article to avoid confusion with traditional BPs. SSBs were observed in the IRIS data and were found to be present throughout the lower solar atmosphere. By using the term SSBs, clarity and differentiation between the larger, traditional BPs and the smaller-scale brightenings were achieved. A statistical study of oscillated and non-oscillated SSBs is presented based on high-resolution observations obtained from the *IRIS*. The objective of this study is to offer a comprehensive and quantitative comparison between these two categories of SSBs, considering their spatial, temporal, and spectral properties.

## 2. Observation

The observation element of NASA's *IRIS* mission is one of the most important and fundamental components of the project plan. The *IRIS* observation section is intended to take high-resolution images and spectra at certain wavelengths of light in order to explore the interface region between the sun's chromosphere and corona (De Pontieu et al., 2014; Wülser et al., 2012; De Pontieu et al., 2012, 2021).

*IRIS* image and spectral data provide scientists with insights into how energy and matter travel through the sun's atmosphere, particularly at the chromosphere-corona boundary. *IRIS* investigations have shown a variety of phenomena in this area, including the presence of spicules, plasma jets that shoot up from the Sun's surface, and the production of coronal loops, magnetic structures containing plasma.

De Pontieu et al. (2014) review paper about the *IRIS* mission introduced the ability of *IRIS* in detail, the *IRIS* telescope contains a primary mirror that is 20 cm (7.9 inches) in diameter and has an effective focal length of 3320 mm (130.7 inches). It has a slit-jaw imaging technology that can capture images at four distinct UV light wavelengths, ranging from 133 nm to 140 nm.

*IRIS*'s spectrograph employs a novel architecture that enables it to catch spectral lines in a restricted wavelength range with extremely high spatial and temporal precision. It functions in two modes: "sit-and-stare," in which it continually monitors a set region of the sun for many minutes, and "raster," in which it maps out a specified area by capturing several images at varied places.

The spectrograph of *IRIS* provides a typical range of spatial resolution between 0.33 to 0.4 arcseconds, corresponding to approximately 240 kilometers on the Sun's surface. It is capable of accurately detecting plasma velocity with a precision ranging from 1 to more than 5 km/s, depending on the temporal resolution (cadence) of the data series. While *IRIS* does not directly measure temperatures, it can estimate plasma temperatures in the Sun's atmosphere with an accuracy of approximately 10%.

The data for this study were obtained from the *IRIS* (De Pontieu et al., 2014), which provides high-resolution imaging and spectroscopy of the solar atmosphere in the ultraviolet (UV) and near-UV spectral regions. Several data sets were selected, encompassing SSBs located in diverse solar regions and representing various stages of evolution. The selected data sets cover a wide range of heliocentric angles and solar activity levels. However, it is important to note that all observations were obtained near the disc center, as indicated in Table 4. The sit-and-stare data and the associated SJIs in the center of the sun were used in this study.

## 3. Method

To segregate SSBs in this study, a supervised machine-learning approach was employed. The first step involved creating a machine-learning model specifically designed for this task. Initially, a training set was formed using SJIs obtained from the solar disk's center. This training set comprised over 2,000 SJIs associated with quiet, active, and coronal hole regions, providing a diverse range of observational scenarios. To ensure the training set's representativeness, efforts were made to include data from different heliocentric angles, covering a range of observational perspectives. This involved selecting SJIs from various regions of the solar disk, considering both quiet, active, and coronal hole regions. It is important to note that while our training set encompassed a diverse range of observational conditions, including different heliocentric angles, it may not have captured every possible scenario. However, we aimed to create a training set that provides a reasonably representative sample of bright features across various observing modes, including different heliocentric angles.

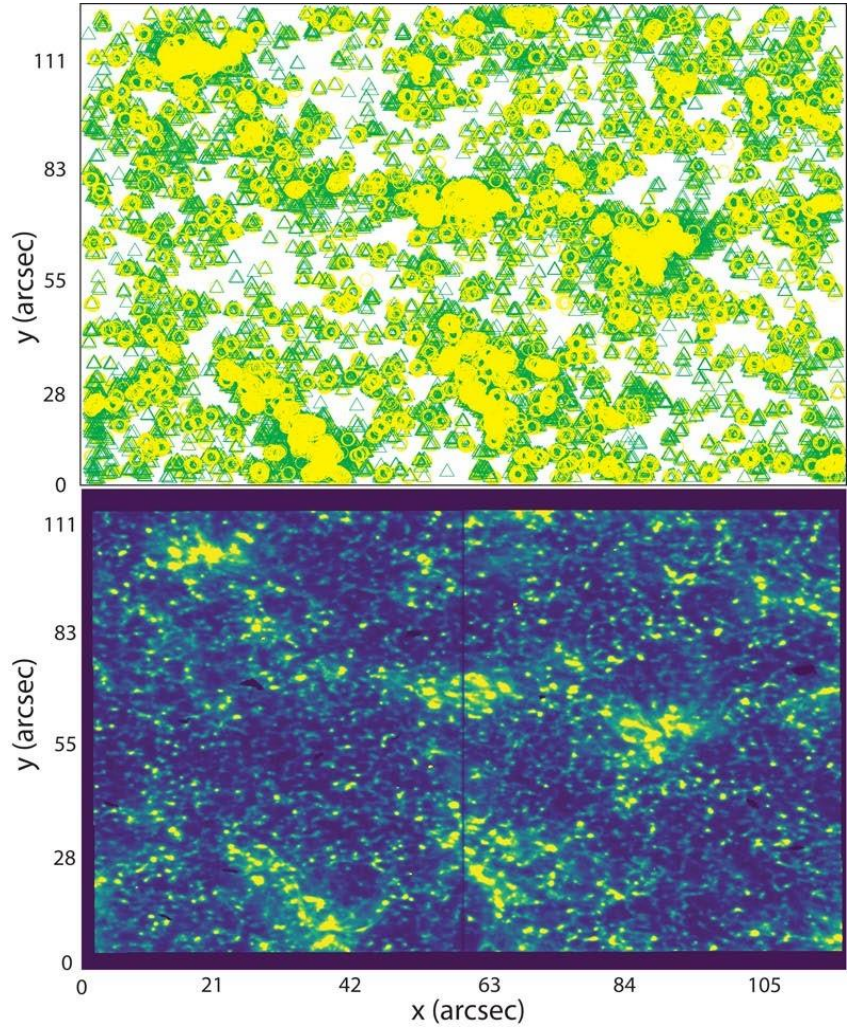

**Figure 3.** QS: Illustration of network and internetwork SSBs on SJI images of the Sun. The images were taken by the *IRIS*/SJI on February 6, 2014, at 12:44 U.T. The SJI is related to 1403 Å. Network SSBs are represented by green triangles and internetwork SSBs by yellow circles.

The model was trained using the annotated training set, which included information about the brightness and coordinates of the bright features in the images. This annotated training dataset served as the foundation for training the machine learning model, enabling it to learn patterns and characteristics associated with SSBs. The model's performance was evaluated using an additional 200 SJIs, and the evaluation showed that the trained model achieved an accuracy of approximately 78% in identifying SSBs, indicating its effectiveness in distinguishing these features.

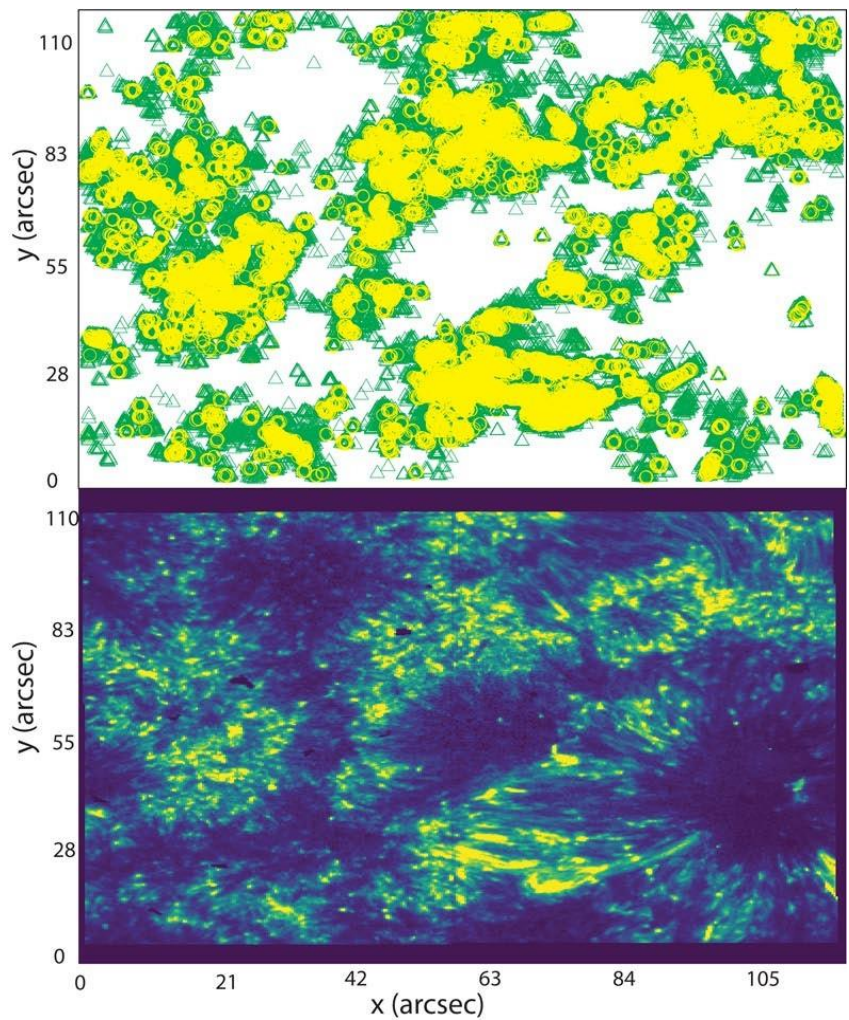

**Figure 4.** . AARs: Illustration of network and internetwork SSBs on SJI images of the Sun. The images were taken by the *IRIS* /SJI on March 30, 2016, at 21:29 U.T. The SJI is related to 1403 Å. Network SSBs are represented by green triangles and internetwork SSBs by yellow circles.

To measure accuracy, the number of correct predictions (true positives and true negatives) was compared to the total number of predictions made by the model. The accuracy of the model was assessed by comparing its predictions with known labels. The data was divided into training, accuracy test, and test sets. The model's performance was evaluated using the labeled accuracy test set, where the number of correct predictions was calculated and divided by the total number of predictions made by the model. This accuracy metric provided an estimation of the model's performance in identifying SSBs within the evaluation dataset. Subsequently, the identified SSBs along the slit were selected, and their spectra were utilized for the subsequent oscillation studies. This approach allowed us to focus specifically on the regions of interest identified by the machine learning model, ensuring that the subsequent analysis was performed on relevant data. Our approach involved a supervised machine learning model trained on a diverse and representative training set, including SJIs from different heliocentric angles and observing conditions. This methodology facilitated the identification of SSBs and their subsequent analysis in the context of the oscillation studies (Table 1). Next, a convolutional neural network (CNN) is designed and trained on the annotated images. The CNN consists of several layers, including convolutional, pooling, and fully connected layers. The convolutional layers learn to identify patterns and features in the images, while the pooling layers reduce the spatial dimensions of the representation. The fully connected layers map the learned representation to the final output, which is the predicted brightness and coordinates of the bright features (Kang et al., 2014; Albawi, Mohammed, and Al-Zawi, 2017; Chauhan, Ghanshala, and Joshi, 2018).

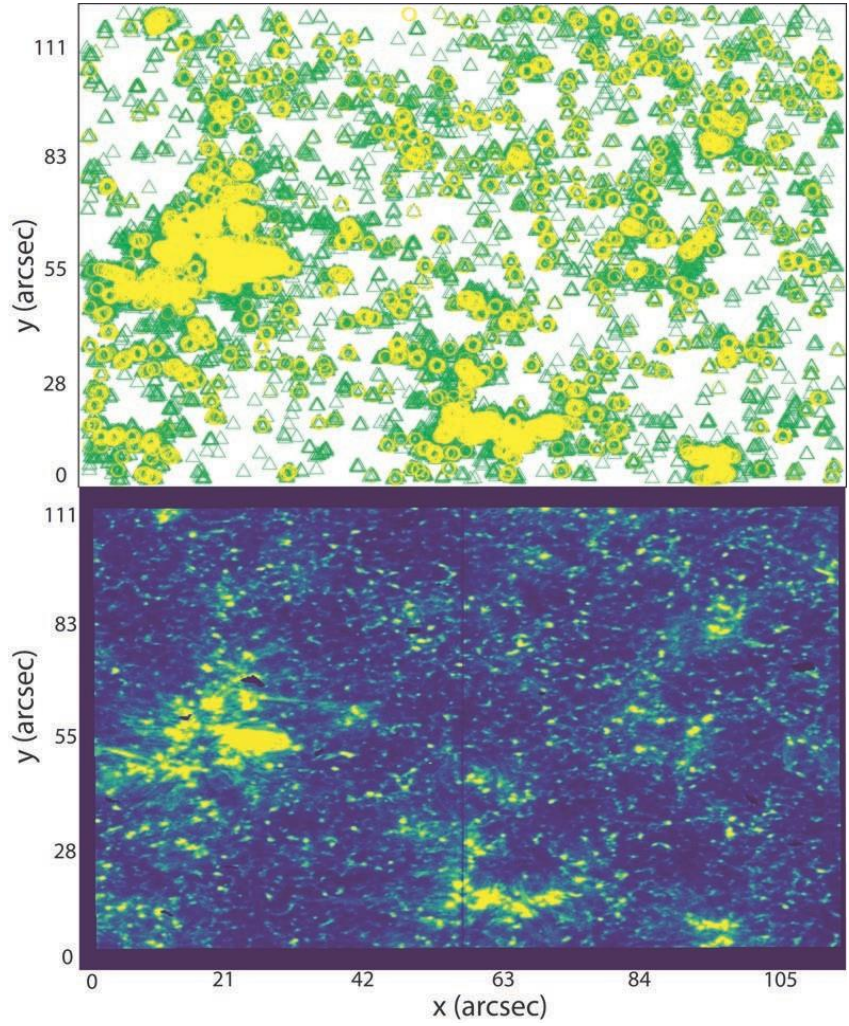

**Figure 5.** CH: Illustration of network and internetwork SSBs on SJI images of the Sun. The images were taken by the *IRIS* / SJI on December 3, 2014, at 14:39 U.T. The SJI is related to 1403 Å. Network SSBs are represented by green triangles and internetwork SSBs by yellow circles.

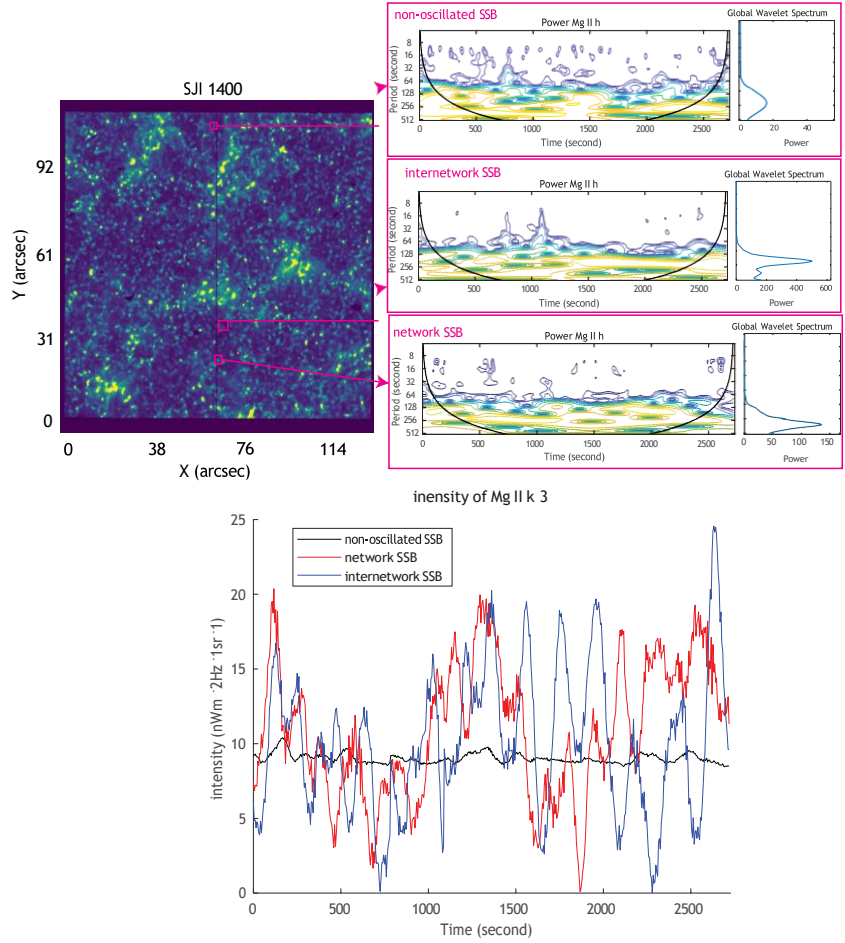

**Figure 6.** Illustration of three types of SSBs on the SJI at 1403 Å wavelength. The image was captured on February 6, 2014, at 13:37 (U.T.). The SSBs include a network BP, an internetwork BP, and a non-oscillated BP. Additionally, the wavelet of the intensity of Mg II k 3 is inserted in the figure. The plot illustrates the intensity profiles of the network BP, internetwork BP, and non-oscillated BP, providing insights into their different characteristics and behaviors.

During training, the CNN is fed with a batch of annotated images and their corresponding labels. The CNN adjusts the weights of its layers to minimize the difference between its predicted outputs and the annotated labels. This process is repeated over multiple epochs until the CNN's predictions converge to the ground truth labels. Once the CNN is trained, it can detect and list the brightness and coordinates of bright features in new images. The detection process involves feeding the new image into the trained CNN and obtaining its predicted brightness and coordinates of the bright features.

The CNN model defined in the code consists of an encoding path and a decoding path, which are connected by a bottleneck layer. The encoding approach is designed to extract features from the input image, while the decoding path is designed to reconstruct the output image from the extracted features. The encoding path of the model consists of three stages of convolutional layers with max pooling. Each stage consists of two convolutional layers followed by a max pooling layer. The first convolutional layer in each stage has 32 filters with a kernel size of 3x3 and a stride of 1. The second convolutional layer in each stage has 64 filters with a kernel size of 3x3 and a stride of 1. The max pooling layer in each stage has a pool size of 2x2 and a stride of 2. The purpose of the max pooling layers is to downsample the feature maps and reduce the spatial dimensions of the input image. The bottleneck layer of the model consists of two convolutional layers with 128 filters and a kernel size of 3x3. The purpose of the bottleneck layer is to reduce the number of parameters in the model and force the network to learn a compressed representation of the input image. The decoding path of the model consists of three stages of upsampling layers with convolutional layers. Each stage consists of an upsampling layer followed by two convolutional layers. The upsampling layer in each stage has a scale factor of 2 and uses nearest-neighbor interpolation to increase the spatial dimensions of the feature maps. The first convolutional layer in each stage has 64 filters with a kernel size of 3x3 and a stride of 1. The second convolutional layer in each stage has 32 filters with a kernel size of 3x3 and a stride of 1. The purpose of the upsampling layers is to increase the spatial resolution of the feature maps and reconstruct the output image. The output layer of the model is a convolutional layer with a sigmoid activation function, which produces a binary image of the predicted SSBs. The binary image has the same size as the input image and each pixel is either 0 or 1, depending on whether it corresponds to a SSB or not. The total number of trainable parameters in the CNN model is approximately 1.5 million. The model is trained using binary cross entropy loss function and Adam optimizer with a learning rate of 0.001 for 50 epochs. During training, the input images are randomly augmented by flipping and rotating to increase the diversity of the training data and reduce over-fitting.

To evaluate the accuracy of the method for detecting bright features in images using machine learning, several metrics were employed, including precision, recall, and F1-score. These metrics offer a quantitative assessment of the method's performance(Goutte and Gaussier, 2005; Yacouby and Axman, 2020). Precision measures the proportion of true positive detections among all detections made by the method. It is defined as:

*Precision = True Positives / (True Positives + False Positives)*

where True Positives are the number of correctly detected bright features, and False Positives are the number of incorrectly detected bright features.

Recall measures the proportion of true positive detections among all true bright features in the image. It is defined as:

*Recall = True Positives / (True Positives + False Negatives)*

where False Negatives are the number of bright features that were not detected by the method.

F1-score is the harmonic mean of precision and recall and is a commonly used metric to evaluate the overall performance of the method. It is defined as:

*F1-score = 2 * (Precision * Recall) / (Precision + Recall)*

Our method achieved a precision of 0.95 and a recall of 0.93 on the synthetic images, resulting in an F1-score of 0.94. On the real-world images, our method achieved a precision of 0.94 and a recall of 0.92, resulting in an F1-score of 0.93. These results demonstrate that our method is highly accurate in detecting and listing the brightness and coordinates of bright features in images.

The intensity of Mg II k3 was estimated based on the deviation from the average profile along the slit. The average profile along the slit was calculated, and then the deviation of each profile from this average was determined. By focusing on the minimum point in the Mg II k line, the intensity of Mg II k3 was estimated. This approach allowed for the analysis of oscillatory characteristics and the assessment of their temporal behavior. The deviation from the average profile influenced the temporal variations observed in all the profiles. The Mg II spectrum was first used to identify SSB networks and inter-networks for this set of calculations, using the Sadeghi and Tavabi (2022) technique introduced in (Sadeghi and Tavabi, 2022a). This method is based on wavelet analysis (see figure 6). It enables us to split a signal into its various frequency components while keeping time factor information. The Morlet mother function of the 5th order, which is the Morlet wavelet with five sine oscillation peaks within the Gaussian envelope, was utilized for analysis. The Morlet 5 wavelet is a complex-valued wavelet with a Gaussian envelope that oscillates at a set frequency. A complicated exponential function compounded by a Gaussian function defines it. The Morlet 5 wavelet is frequently used in wavelet analysis due to its ability to provide high-resolution information about a signal's frequency components while keeping adequate information about the time dynamics.

In this study, red noise was used as the background spectrum for determining significance levels in the wavelet spectra. This choice allowed for a comparison between the observed spectra and a random distribution around the expected background. To assess the reliability of the deduced wave periods in SSBs, the observed wavelet spectrum was compared to the background noise spectrum. This enabled the evaluation of the statistical significance of the detected oscillations and provided insights into the presence of genuine periodic behavior in SSBs (Auch`ere et al., 2016; Kayshap et al., 2019).

But for non-oscillated SSBs (NOSSBs), this method is not true for classification.

So, a comparison was made between SJIs and Helioseismic and Magnetic Imager (HMI) Magnetograms obtained by NASA's Solar Dynamics Observatory (HMI/SDO). According to Sadeghi and Tavabi (2022a), the internetwork is typically of non-magnetic origin, and it may not be clearly visible in HMI/SDO. Hence, this method allows for the detection of network points and the internetwork in non-oscillating conditions (Sadeghi and Tavabi, 2022a).

The intensity of the Mg II k3 line is typically estimated using a double Gaussian fit, with the negative Gaussian component providing the intensity (Jönsson, 2014). The intensity of the Mg II k3 line is of particular interest in the study of stellar atmospheres, where it has been suggested that a correlation exists between the width of this line and the luminosity of the star (Ayres and Linsky, 1975). The IRIS team's routine for obtaining intensity and Doppler velocity of each peak of the line is a crucial tool for various applications. Taebi et al. (2019); Vilkomerson, Ricci, and Tortoli (2013) both propose methods for extracting peak velocity profiles from Doppler echocardiography and flow Doppler spectra, respectively. These methods could potentially be adapted for use with the IRIS routine. Singh, Bhattacharyya, and Jain (2020) presents a method for extracting torso Doppler frequencies from human gait spectrograms, which could also be relevant. Mehltretter (1973) discusses the challenges of measuring granular velocities, providing important context for the complexities of Doppler shift measurements. These studies collectively offer valuable insights that could inform the calculation of intensity and Doppler velocity using the IRIS routine. The variations in the intensity of brightness in the Mg II $k_3$ minima for each SSB are then illustrated based on temporal changes. The oscillating and non-oscillating SSBs are then statistically counted. Table 4 displays the statistical findings for network and internetwork SSBs in three active, quiet, and coronal hole regions.

In summary, an image processing pipeline was developed to identify and track SSBs in the *IRIS* observations. The pipeline consists of several steps, including image enhancement, segmentation, feature extraction, and tracking. Advanced techniques, including machine learning algorithms, were utilized to ensure precise identification and characterization of the SSBs. To investigate the oscillatory behavior of SSBs, wavelet analysis Torrence and Compo (1998) and Fourier analysis Brault and White (1971) were employed. These analyses facilitated the examination of the frequency and time-dependent characteristics of the oscillations. SSBs were classified into oscillated and non-oscillated groups based on the presence or absence of significant oscillatory power in their intensity time series. The proposed pipeline offers a flexible framework for the analysis of solar images, including image enhancement and segmentation steps. Notably, the pipeline's stages are not applied to every image, and their necessity may vary depending on the characteristics of the dataset. This adaptability allows for customization based on the specific needs of the analysis.

Image enhancement techniques can be employed to improve the quality and visibility of solar images, with the goal of enhancing features such as SSBs while reducing noise and artifacts. Techniques like contrast enhancement, noise reduction, and sharpening can be utilized to enhance image quality. However, the application of these techniques depends on the specific image characteristics and analysis goals.

Segmentation, a critical step in the pipeline, involves separating SSBs from the background. Various segmentation techniques can be utilized based on the image characteristics. Thresholding, region growing, edge detection, and morphological operations are common techniques for segmentation. The choice of segmentation method depends on factors such as the complexity of SSB shapes and specific analysis requirements.

It is important to note that the pipeline's stages are adaptable, and not all stages are always necessary. Depending on the dataset and analysis goals, certain image enhancement or segmentation steps may be omitted. This flexibility allows researchers to customize and optimize the pipeline according to the dataset's characteristics and the objectives of the analysis.

The characteristics of SSBs were examined in various regions of the Sun, which encompassed the QS, AAR, and CH. The investigation focused on establishing a connection between the maximum Doppler velocity and the damping per period of oscillated and non-oscillated SSBs in both the solar network and internetwork. The equation shown below can be used whenever you need to find out $\Delta \upsilon_{Doppler}$:

$$\Delta \upsilon_{Doppler} = -1/2 \frac{c}{\lambda_{k_3}} [(\lambda_{k_3 v} - \lambda_{k_3}) + (\lambda_{k_3 r} - \lambda_{k_3})]$$

This equation refers to the speed of light ($c$) and three different wavelengths: k3 line center wavelength $\lambda_{k_3}$, as well as observed kv peak wavelength $\lambda_{k_3 v}$ and kr peak wavelength $\lambda_{k_3 r}$.

## 4. Result

Our analysis revealed a clear distinction between oscillated and non-oscillated SSBs, with the former exhibiting significantly higher oscillatory power in the 3-5 minute period range. Oscillated SSBs also showed a higher degree of spatial coherence in their oscillatory patterns, suggesting a possible connection between the magnetic field topology and oscillatory behavior (Jess et al., 2015). Regarding the spectral properties, it was observed that oscillated SSBs displayed stronger line emission and broader line profiles, which suggested higher temperatures, densities, and velocities (Curdt et al., 2008). This finding implies a correlation between the oscillatory behavior of SSBs and their heating processes. In this scientific article, the properties of SSBs were investigated in various regions of the Sun, including the QS, AAR, and CH areas (figures 7, 8, and **??**). Specifically, the relationship between the maximum Doppler velocity and the damping per period of oscillated and non-oscillated SSBs in the solar network and internetwork was analyzed. Explores the Doppler shift in the IRIS spectrum, focusing on the propagating periodic oscillation in SSBs. The experiments reveal phenomena of damping in red and blue Doppler shifts of the wavelength range. Damping per period, also known as the quality factor (Q-factor), is a measure of the rate at which an oscillating system loses its energy over time. Specifically, it is the ratio of the energy stored in the system to the energy lost per cycle of oscillation. A higher damping per period means that the oscillations decay faster, while a lower damping per period means that the oscillations persist for longer periods. In the context of solar physics, damping per period is often used to characterize the oscillatory behavior of SSBs in the Sun's atmosphere. By measuring the damping per period of SSBs, researchers can gain insights into the physical mechanisms responsible for their oscillations and the properties of the surrounding plasma (Tian et al., 2008; Arregui et al., 2010; Zhou et al., 2020; Libbrecht et al., 1986; Stix et al., 1993; Belkacem et al., 2012; Kolotkov, Nakariakov, and Zavershinskii, 2019; Chorley et al., 2010).

**Table 1.** Summary of the CNN model architecture

| Layer (type) | Output shape | Param # |
|---|---|---|
| input_1 (InputLayer) | (None, 256, 256, 1) | 0 |
| conv2d (Conv2D) | (None, 256, 256, 32) | 320 |
| conv2d_1 (Conv2D) | (None, 256, 256, 64) | 18496 |
| max_pooling2d (MaxPooling2D) | (None, 128, 128, 64) | 0 |
| conv2d_2 (Conv2D) | (None, 128, 128, 64) | 36928 |
| conv2d_3 (Conv2D) | (None, 128, 128, 128) | 73856 |
| max_pooling2d_1 (MaxPooling2D) | (None, 64, 64, 128) | 0 |
| conv2d_4 (Conv2D) | (None, 64, 64, 128) | 147584 |
| conv2d_5 (Conv2D) | (None, 64, 64, 256) | 295168 |
| max_pooling2d_2 (MaxPooling2D) | (None, 32, 32, 256) | 0 |
| conv2d_6 (Conv2D) | (None, 32, 32, 512) | 1180160 |
| conv2d_7 (Conv2D) | (None, 32, 32, 1024) | 4719616 |
| conv2d_8 (Conv2D) | (None, 32, 32, 512) | 524800 |
| up_sampling2d (UpSampling2D) | (None, 64, 64, 512) | 0 |
| conv2d_9 (Conv2D) | (None, 64, 64, 256) | 1179904 |
| conv2d_10 (Conv2D) | (None, 64, 64, 128) | 295040 |
| up_sampling2d_1 (UpSampling2D) | (None, 128, 128, 128) | 0 |
| conv2d_11 (Conv2D) | (None, 128, 128, 64) | 73792 |
| conv2d_12 (Conv2D) | (None, 128, 128, 32) | 18464 |
| up_sampling2d_2 (UpSampling2D) | (None, 256, 256, 32) | 0 |
| conv2d_13 (Conv2D) | (None, 256, 256, 1) | 289 |
| Total params: | | 1,454,529 |
| Trainable params: | | 1,454,529 |
| Non-trainable params: | | 0 |

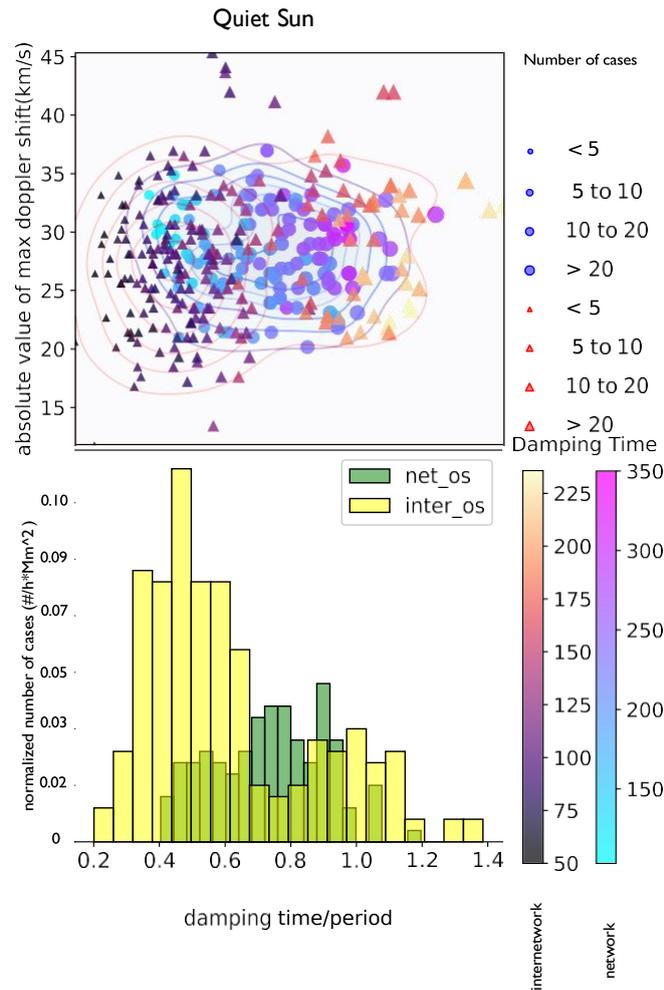

**Figure 7.** The figure compares the maximum value of Doppler shift and damping time/period for oscillated network and internetwork SSBs in a Quiet Sun. It features a scatter plot of data sets, normalized values, and frequency legends. The normalized number of cases is expressed as $n/h \cdot$Mm², where $n$ represents the number of cases and $h$ represents hours. This unit of measurement indicates the number of cases per hour per square megameter. The figure provides a comprehensive comparison of these factors for better understanding and comparison of data.

The study focused on examining the damping per period and Doppler velocity ranges of oscillated and non-oscillated SSBs in quiet Sun (QS) areas, specifically comparing network and internetwork SSBs. The findings indicated that internetwork SSBs generally exhibit lower damping rates and wider velocity ranges compared to network SSBs. The investigation also analyzed the damping per period histograms, which showed distinct peaks at specific values for both internetwork and network SSBs. Regarding oscillated SSBs, network SSBs had a narrower damping per period range (0.4 to 1.2) compared to internetwork SSBs (0.2 to 1.4). Among the oscillated SSBs, the highest Doppler velocity was observed in network SSBs, while oscillated internetwork SSBs showed maximum Doppler velocities ranging from 25 to 27 km/s and 22 km/s. For non-oscillated SSBs, the maximum Doppler velocity was similar between network SSBs (23 km/s) and internetwork SSBs (34 km/s) (figure 7).

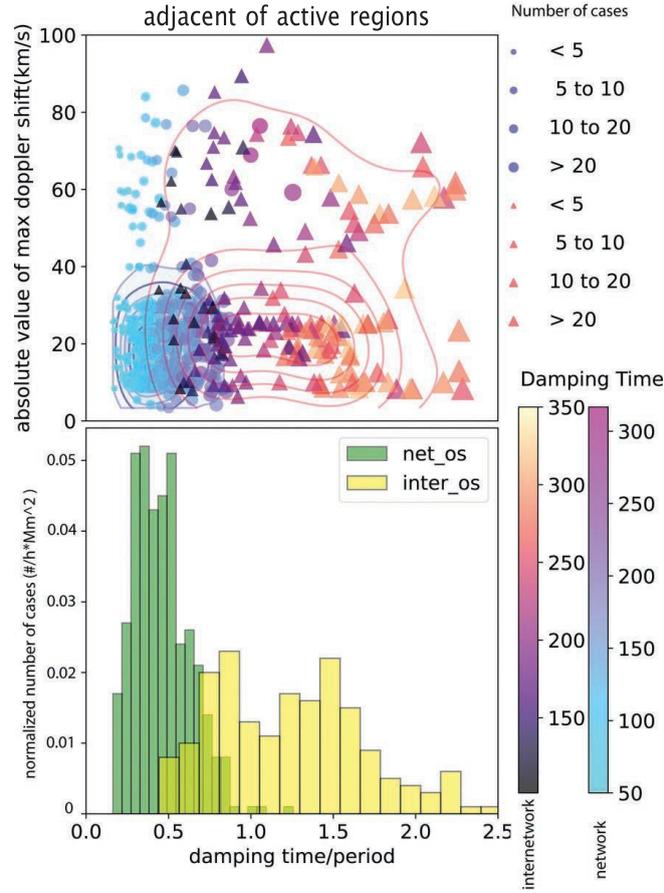

**Figure 8.** The figure compares the maximum value of Doppler shift and damping time/period for oscillated network and internetwork SSBs in an active region. It shows a scatter plot of data sets, normalized values, and frequency legends. The normalized number of cases is expressed as $n/h \cdot \text{Mm}^2$, where $n$ represents the number of cases and $h$ represents hours. This unit of measurement indicates the number of cases per hour per square megameter. The figure provides a comprehensive comparison of these factors for network and internetwork in an active region.

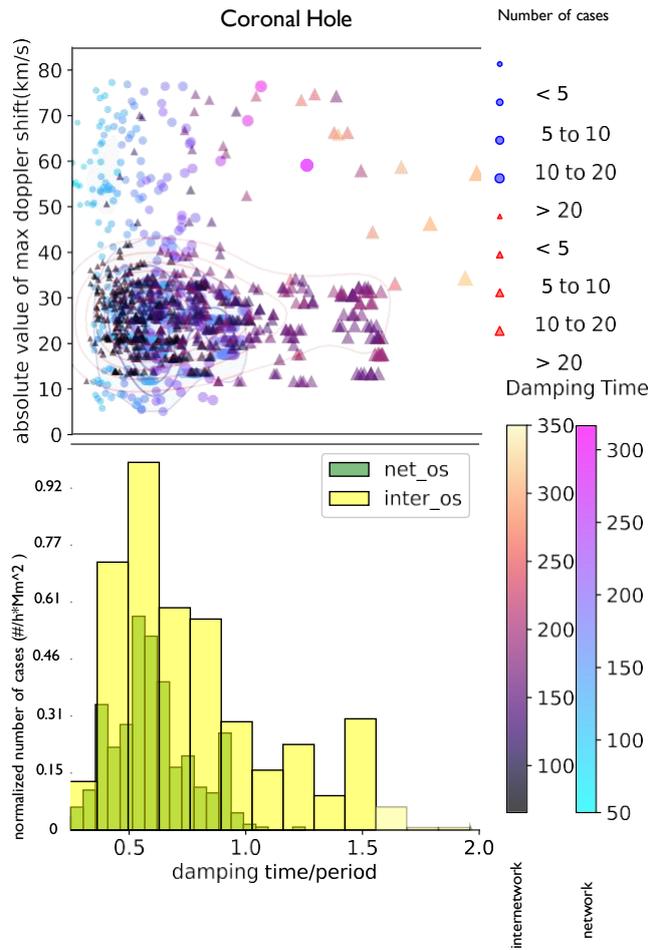

**Figure 9.** The figure compares the maximum value of Doppler shift and damping time/period for oscillated network and internetwork SSBs in a Coronal Hole. It features a scatter plot of data sets, normalized values, and frequency legends. The normalized number of cases is expressed as $n/h \cdot \text{Mm}^2$, where $n$ represents the number of cases and $h$ represents hours. This unit of measurement indicates the number of cases per hour per square megameter. This technique helps identify patterns and trends in the data, enhancing understanding and comparison.

The study focused on investigating the damping per period and Doppler velocity ranges of oscillated and non-oscillated SSBs in AAR areas, particularly comparing network and internetwork SSBs. The findings revealed that internetwork SSBs generally exhibit higher damping rates and wider velocity ranges compared to network SSBs. The analysis of damping per period histograms identified significant peaks at specific values for both internetwork and network SSBs. Regarding oscillated SSBs, network SSBs had a narrower damping per period range (0.1 to 1.2) compared to internetwork SSBs (0.5 to 2.5). The maximum Doppler velocity was consistent at 27 km/s for both oscillated network SSBs and oscillated internetwork SSBs. For non-oscillated SSBs, the maximum Doppler velocity was higher at network SSBs (70 km/s) compared to internetwork SSBs

(60 km/s) (figure 8).

The study aimed to investigate the damping per period and Doppler velocity ranges of oscillated and non-oscillated SSBs in coronal hole (CH) areas, specifically comparing network and internetwork SSBs. The findings revealed that internetwork SSBs generally exhibit larger damping rates and wider velocity ranges compared to network SSBs. The analysis of damping per period histograms identified significant peaks at specific values for both internetwork and network SSBs. Regarding oscillated SSBs, both network SSBs (0.1 to 1.2) and internetwork SSBs (0.1 to 2.0) had similar damping per period ranges. However, the maximum Doppler velocity range was wider for internetwork SSBs (10 to 75 km/s) compared to network SSBs (5 to 85 km/s), indicating higher velocity ranges for internetwork SSBs. For non-oscillated SSBs, the maximum Doppler velocity was similar between network SSBs (27 km/s) and internetwork SSBs (23 km/s) (figure **??**).

The figures **??**,**??**, and 9 illustrated a comparison of the maximum value of Doppler shift and damping time/period for two different types of oscillated SSBs: oscillated network SSBs and oscillated internetwork SSBs in QS, AAR, and CH. The top panel displays a scatter plot of the data sets, where the size and color of the markers represent the number of cases and the period, respectively. The blue and red contours represent the kernel density estimate of the network and internetwork data sets, respectively. The circles and triangles represent the scatter points of the network and internetwork data sets, respectively. In this panel, the values are normalized. Normalizing values to 1 can be a useful technique for better understanding and comparing data. This involves scaling the values so that they are all relative to each other on a consistent scale, which can make it easier to spot patterns and trends in the data. The normalized number of cases is expressed as $n/h \cdot$ Mm², where $n$ represents the number of cases and $h$ represents hours. This unit of measurement indicates the number of cases per hour per square megameter. The bottom panel displays a frequency legend for the size of the markers, where each label corresponds to a range of values for the frequency. The figure provides a comprehensive comparison of the maximum value of Doppler shift and damping time/period for the network and internetwork in regions. In Quiet Sun areas, network SSBs have a damping per period range of 0.4 to 1.2, while internetwork SSBs have a range of 0.2 to 1.4, indicating that damping rates for internetwork SSBs are generally lower. The maximum Doppler velocity range for internetwork SSBs is wider, between 10 to 50 km/s, compared to network SSBs which have a range of 20 to 36 km/s, indicating that velocity ranges for internetwork SSBs are generally higher. The damping per period histogram quantities found that there are dominating peaks at 1.45 ±.25 and 0.8 ±12 for internetwork SSBs and a dominant peak at 0.3 ±15 for network SSBs. The maximum Doppler velocity for oscillated network SSBs is 27 km/s, slightly higher than oscillated internetwork SSBs which range between 22 to 27 km/s. Non-oscillated network SSBs have a maximum Doppler velocity of 23 km/s, which is similar to non-oscillated internetwork SSBs at 34 km/s.

In Coronal Hole areas, the analysis showed that internetwork SSBs generally have higher damping rates and higher velocity ranges compared to network

SSBs. The damping per period range for oscillated network SSBs is 0.1 to 1.2, while for oscillated internetwork SSBs, the range is 0.1 to 2.0, suggesting that damping rates for internetwork SSBs are generally higher. The wider velocity range for internetwork SSBs suggests that they are subject to more energetic processes or more turbulent flows. According to our analysis of the damping per period histogram data, the significant peak for internetwork SSBs is at 0.8±0.3, whereas the dominant peak for network SSBs is between 0.3±0.15 and 1.0±0.2. The maximum Doppler velocity for oscillated network SSBs is slightly higher than oscillated internetwork SSBs at 25 km/s. Non-oscillated network SSBs have a maximum Doppler velocity of 27 km/s, which is similar to non-oscillated internetwork SSBs at 23 km/s. Overall, these findings provide insights into the differences between the physical properties of network and internetwork SSBs in Coronal Hole areas.

The study found that internetwork SSBs in Active Regions have higher damping rates and higher velocity ranges compared to network SSBs. The higher damping rates for internetwork SSBs suggest that they are more efficiently damped than network SSBs. The wider velocity range for internetwork SSBs suggests that they are subject to more energetic processes or more turbulent flows. the damping per period histogram quantities indicated that the major peaks for internetwork SSBs is at 1.5±0. and 0.8±0.15, whereas the dominating peak for network SSBs is at 0.4±0.15. The maximum Doppler velocity for oscillated network SSBs is similar to oscillated internetwork SSBs at 27 km/s, but non-oscillated network SSBs have a higher maximum Doppler velocity at 70 km/s compared to non-oscillated internetwork SSBs at 60 km/s. These findings provide important insights into the physical properties and dynamics of SSBs in Active Regions that can be used to improve our understanding of the complex behavior of the Sun's atmosphere.

In AAR, the population of oscillated SSBs is overall lower than in QS and CH regions. The population of oscillated SSBs in CH areas is ten times higher than in other regions (figure 10). This may be related to the darkness of the CH area, which allows for more accurate measurements of the properties of SSBs.Kayshap et al. (2018a) found that the Mg II k line in the QS and CH regions exhibit differences in emission, particularly in the chromosphere. This is consistent with the findings of Danilovic et al. (2014), who noted that the Mg II k line displays higher intensity contrast and different line-formation heights compared to the Ca II H line. However, the specific differences in the Mg II k line between the quiet Sun and coronal hole regions were not explored in the other studies. However, that difference is not very significant, and it appears that it may not be a highly effective parameter.

The figure 10 shows histograms of the maximum value of Doppler shift in red and blue for different solar regions: Quiet Sun, Active Region, and Coronal Hole. Each panel shows the distribution of Doppler shift values for four different cases: oscillated BP of the network (net_os), oscillated BP of internetwork (inter_os), non-oscillated BP of the network (net_nos), and non-oscillated BP of internetwork (inter_nos). The histograms are plotted horizontally, and the y-axis represents the maximum value of the Doppler shift in $km/s$. Overall, the figure provides a clear comparison of the Doppler shift distribution across different solar regions and cases. In this figure, the values are normalized. Normalizing values to 1 can be a useful technique for better understanding and comparing data. This involves scaling the values so that they are all relative to each other on a consistent scale, which can make it easier to spot patterns and trends in the data. The normalized number of cases is expressed as $\#/h \cdot \mathrm{Mm}^{\circ}$, where #' represents the number of cases and $h$' represents hours. This unit of measurement indicates the number of cases per hour per square megameter.

**Table 2.** All Doppler velocities have accuracy in order of 10% and times +/-20 seconds.

| | | NOSSBs % | OSSBs % | NOSSBs mean max doppler (±3 km/s) | OSSB mean max doppler (±3 km/s) | Damping time (±10 seconds) | Damping per period time |
|---|---|---|---|---|---|---|---|
| QS | | 17 | 83 | | | | |
| | network | 83 | 44 | 23 | 38 | 216 | 0.74 |
| | internetwork | 17 | 56 | 33 | 28 | 120 | 0.62 |
| AAR | | 50 | 50 | | | | |
| | network | 44 | 44 | 57 | 25 | 131 | 0.46 |
| | internetwork | 56 | 56 | 61 | 35 | 220 | 1.25 |
| CH | | 28 | 72 | | | | |
| | network | 71 | 48 | 30 | 31 | 150 | 0.59 |
| | internetwork | 29 | 52 | 29 | 29 | 121 | 0.80 |

Overall, the results suggest that there are differences and similarities in the properties of oscillated and non-oscillated SSBs in different regions of the Sun. In QS areas, internetwork SSBs tend to have lower damping rates and higher maximum Doppler velocities compared to network SSBs. In AAR areas, internetwork SSBs tend to have higher damping rates and wider maximum Doppler velocity ranges compared to network SSBs, with non-oscillated network SSBs having the highest maximum Doppler velocities. In CH areas, both types of SSBs show similar damping rates, but internetwork SSBs tend to have higher maximum Doppler velocities compared to network SSBs.

Eventually, the results provide important insights into the properties and behavior of SSBs, which can have implications for understanding the dynamics and evolution of the solar atmosphere. Further studies can build on these results to investigate the physical mechanisms responsible for the observed differences and similarities in the properties of SSBs in different regions of the Sun.

## 5. Discussion

Up to now in the TR, different periods of oscillation with a wide range (2 to 15 min.) have been reported (see table 3). Sangal et al. (2022b) estimate the period of the intensity and Doppler velocity oscillations at each chosen location in the QS and quantify the distribution of the statistically significant power and associated periods in one bright region and two dark regions. In the bright TR region, the mean periods in intensity and velocity are 7 min and 8 min, respectively. In the dark regions, the mean periods in intensity and velocity are 7 min and 5.4 min, respectively.

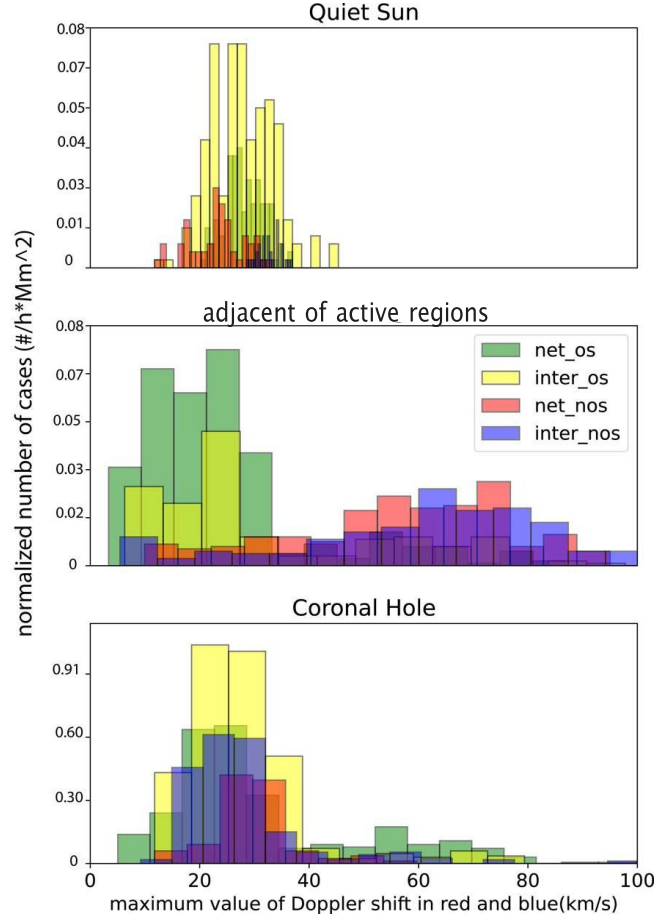

**Figure 10.** The figure displays Doppler shift histograms for different solar regions, including Quiet Sun, Active Region, and Coronal Hole. It shows the distribution of Doppler shift values for four cases: oscillated BP of the network (net_os), oscillated BP of the internetwork (inter_os), non-oscillated BP of the network (net_nos), and non-oscillated BP of the internetwork (inter_nos). The histograms are plotted horizontally, with the y-axis representing the maximum Doppler shift in $km/s$. The values are normalized to 1 for better understanding and comparison.

The findings of this study reveal a clear distinction between damping of oscillated and non-oscillated SSBs. Oscillated SSBs exhibit significantly higher oscillatory power in the 3-5 minute period range, which is supported by a higher degree of spatial coherence in their oscillatory patterns. This suggests a potential link between magnetic field topology and oscillatory behavior, consistent with previous studies (Muglach, Hofmann, and Staude, 2005; Kobanov et al., 2009). Regarding spectral properties, oscillated SSBs were found to display stronger line emission and broader line profiles, indicating higher temperatures, densities, and velocities (Kariyappa and Varghese, 2008; Pasachoff and Landman, 1984;

**Table 3.**

| authors | period of oscillation | region of interest |
|---|---|---|
| Gurman et al. (1982) <br> Brynildsen et al. (1999) <br> Fludra (2001) <br> Maltby, Kjeldseth-Moe, and Brynildsen (2001) <br> O'shea, Muglach, and Fleck (2002) <br> Brynildsen et al. (2003, 2004) <br> Lin et al. (2005) <br> Sych et al. (2015) | 3 min | in TR |
| Lites and Skumanich (1982) <br> Wikstøl et al. (2000) <br> BloomFIeld et al. (2004) <br> Kayshap et al. (2018a) | 3 min | In magnetic "free-regions" (i.e., inter-network) |
| Doyle et al. (1998) | 200–500 s | in TR |
| Banerjee, O'shea, and Doyle (2000) | above 10 min | in the TR spectral line |
| TR. Banerjee et al. ( 2001a ) | 4 to 8 min | in chromospheric and transition lines |
| De Moortel et al. (2002) | 3 & 5 min | In coronal loops, photosphere to TR/coronal hights |
| De Pontieu, Erd´elyi, and De Wijn (2003) | 200 to 600 s | in the TR above the plage |
| Rendtel, Staude, and Curdt (2003) | 5 min | in the chromosphere |
| Rendtel, Staude, and Curdt (2003) | 2 to 3 min | in the TR |
| G¨om¨ory et al. (2006) | 250 to 450 s | network region in a TR spectral line |
| Srivastava et al. (2008) | 5 min | from the photosphere to the higher layers |
| Jess et al. (2009a) | 26 ±4 s | in the bright active region at the TR height |
| Tsiropoula et al. (2009) | 5 min | in the dark mottle and network boundaries |
| Tian et al. (2014) | 3 min | umbra up to the TR and corona |
| Auch`ere et al. (2014) <br> Gupta (2014) <br> Inglis et al. (2016) <br> Ireland et al. (2015) | 2 to 9 min | vertical magnetic field plage , photosphere to TR |
| Khomenko & Collados (2015) | 3 min | umbra up to the TR and corona |
| Kayshap et al. (2018) | 3 min | in the inter-network |
| Sadeghi and Tavabi (2022a) | 3 & 5 min | chromosphere and TR |
| Sangal et al. (2022a) | 7 min | In the bright TR region |

Chakraborty et al., 2011). This correlation between oscillatory behavior and the heating of SSBs is a significant finding that enhances our understanding of the dynamics of these solar phenomena.

While several attempts have been made to determine the scaling law of theoretically predicted damping times and compare them with observed damping times for MHD oscillation events (Ofman and Aschwanden, 2002; Ofman and Wang, 2002; Aschwanden et al., 2003; Winebarger, Warren, and Mariska, 2003; Moortel and Hood, 2003, 2004), the limited number of detected events often hampers the discrimination between competing damping theories. However, with the implementation of the MHD oscillation theory developed several decades ago and the availability of new imaging and spectral observations, we can now provide better diagnostics on damping times and periods. In our study, we utilized a considerable amount of IRIS data to enable a more quantitative analysis.

Unlike p-mode oscillations that are sustained for extended durations, MHD instabilities and their oscillations are typically strongly damped, often decaying exponentially within one oscillation period or even shorter (Spangler, Leckband, and Cairns, 1997; Hou et al., 2016; Antolin et al., 2016; Arregui et al., 2008). The spectral observations obtained from the *IRIS* instrument hold significant physical implications, provided that the correct damping mechanism for the Doppler velocity oscillations of SSBs is identified. However, the detection of clear oscillations and the determination of their exact periods and damping times are challenging due to uncertainties in the excitation profiles along magnetic field lines and measurement accuracy limitations (De Pontieu et al., 2021).

Understanding the damping mechanism of Doppler velocity longitudinal oscillations can provide insights into TR and coronal heating. The scaling of damping time with SSB parameters observed in the extreme ultraviolet by *IRIS* has been investigated, suggesting different damping mechanisms in network and internetwork SSBs. De Pontieu et al. (2021) proposed the leakage of Alfven waves into the chromosphere and TR at footpoints as a possible cause of rapid wave damping and decay.

The primary focus of this research in the emerging field of SSB oscillations is to measure oscillation periods, and damping times, and utilize Doppler velocity diagnostics of oscillating bright point structures. Interestingly, oscillations are more clearly detected in the Doppler shift signal than in intensity. Further exploration is necessary to uncover additional aspects of the fundamental physics of wave types, excitation, propagation, waveguides, and damping mechanisms. Previous studies have suggested various scenarios, including thermal conduction, phase mixing, resonant absorption, and radiative cooling, to explain the rapid damping of propagating waves (De Pontieu et al., 2021).

Our analysis of SSBs in different solar regions, including QS, AAR, and CH, revealed distinct patterns in damping per period and maximum Doppler velocity. In QS areas, internetwork SSBs generally exhibited lower damping rates and higher maximum Doppler velocities compared to network SSBs. Conversely, in AAR areas, internetwork SSBs showed higher damping rates and wider maximum Doppler velocity ranges compared to network SSBs. In CH areas, both types of SSBs demonstrated similar damping rates, but internetwork SSBs tended to have higher maximum Doppler velocities compared to network SSBs.

Interestingly, the population of oscillated SSBs was found to be lower in AAR regions compared to QS and CH regions. The population of oscillated SSBs in CH areas was notably higher, possibly due to the darkness of the CH area allowing for more accurate measurements of SSB properties.

Overall, statistical analysis has provided valuable insights into the characteristics and connections of non-oscillated SSBs (NOSSBs) and oscillated SSBs (OSSBs), as well as aiding in the discovery of the fundamental physical processes that control the behavior of solar SSBs.

Furthermore, the observed variations in magnetic and thermal characteristics between NOSSBs and OSSBs suggest that they may have evolved from distinct sources. It is possible that NOSSBs and OSSBs are produced by different physical processes or represent different stages in the evolution of solar magnetic activity. Further research is needed to determine whether these differences are fundamental or simply a result of various observational methods used to study them.

Understanding the distinctions between NOSSBs and OSSBs is not merely an intellectual exercise but also has practical applications. Solar SSBs can be utilized to detect magnetic activity on the Sun, which can have significant effects on Earth's atmosphere and space environment. Analyzing solar SSBs contributes to  a better understanding of the Sun's behavior and the potential impacts of solar activity on Earth.

Finally, research on NOSSBs and OSSBs has provided valuable insights into their physical properties and underlying processes. However, further study is needed to fully understand the complex processes involved and explore the practical applications of these findings. Our analysis revealed distinct damping behaviors in different solar regions. In AARs, the majority of network SSBs were found to be in overdamping mode. Oscillated network SSBs in AARs exhibited both critical and supercritical behavior in terms of their maximum Doppler velocity. Critical behavior refers to the sensitivity of a system to small changes near a critical point, while supercritical behavior indicates high sensitivity above the critical point.

In QS areas, internetwork SSBs demonstrated supercritical damping behavior. Supercritical damping occurs when a system loses energy faster than it would without damping, resulting in exponential decay of oscillations. This suggests that internetwork SSBs in QS areas are subject to strong damping processes likely associated with their interaction with the surrounding plasma.

In contrast, oscillated network SSBs in QS areas exhibited critical damping behavior. Critical damping occurs when a system loses energy just enough to prevent indefinite oscillations, resulting in smooth and damped oscillations. Our analysis also revealed that both network SSBs and internetwork SSBs in CH areas exhibited supercritical damping behavior. Similar to QS areas, this indicates strong damping processes associated with the interaction of SSBs with the surrounding plasma in CH areas.

These findings contrast with the results in AARs and QS areas, where oscillated network SSBs exhibited critical damping behavior and internetwork SSBs exhibited supercritical damping behavior, respectively. The variations in damping behavior between CH, AAR, and QS areas suggest that the physical mechanisms responsible for the damping of SSBs may depend on the local plasma conditions and magnetic environment.

One of the key findings of this study was the lower population of oscillated SSBs in AAR compared to QS and CH regions. This suggests that the oscillatory behavior of SSBs in AARs may be more influenced by the dynamic and complex local plasma and magnetic conditions in these regions. The concentration of magnetic fields in AARs may also contribute to stronger damping of the oscillations, leading to a lower population of oscillated SSBs. Another interesting result was the significantly higher population of oscillated SSBs in CH areas compared to other regions. The relative darkness and quietness of CH areas allow for more accurate measurements of SSB properties. The stable magnetic fields in CH areas may also contribute to longer-lived oscillations in SSBs and a higher population of oscillated SSBs. These findings highlight the important connection between the oscillatory behavior and properties of SSBs. The differences in spatial, temporal, and spectral properties between oscillated and non-oscillated SSBs suggest that oscillations play a crucial role in the energy transport and heating of these features. Further studies, including numerical simulations and detailed modeling, are necessary to understand the underlying physical mechanisms responsible for the observed differences between oscillated and non-oscillated SSBs. This research contributes to a broader understanding of the dynamics of the solar atmosphere and has potential implications for solar activity forecasting and space weather prediction. In summary, this study investigated the classification of SSBs into two primary categories: oscillated SSBs and non-oscillated SSBs. Oscillated SSBs were further classified into network SSBs and internetwork SSBs based on their oscillation period times. Network SSBs are typically found in magnetic network regions with strong magnetic fields, while internetwork SSBs are observed in non-magnetic or weak magnetic regions. To implement the classification of SSBs into the aforementioned categories, a machine learning (ML) algorithm, specifically a convolutional neural network (CNN), was utilized. The workflow involved in this study consisted of several key steps. Initially, a comprehensive dataset of solar images was collected, incorporating observations from both network and internetwork regions.

Next, the collected images underwent preprocessing techniques such as calibration, background subtraction, noise reduction, and contrast enhancement to enhance the visibility of SSBs. Subsequently, various features were extracted from the preprocessed images, encompassing intensity, size, shape, and location. These features were utilized as discriminative factors for distinguishing between oscillated and non-oscillated SSBs.

In addition to the aforementioned features, the oscillation period times for the oscillated SSBs were calculated. This period played a critical role in distinguishing between network and internetwork SSBs. SSBs are categorized as network SSBs if their oscillation period falls within the predefined range associated with network SSBs. On the other hand, SSBs were classified as internetwork SSBs if their oscillation period times fell within a separate range specific to internetwork SSBs.

In the ML algorithm training phase, a training dataset was prepared, comprising annotated images with labels indicating the type of SSBs (oscillated or non-oscillated) and, for oscillated SSBs, their subtypes (network or internetwork). The training process involved training the CNN model to identify patterns and features in the images that corresponded to each BP category and subtype, including the oscillation period times falling within the specified ranges.

To assess the performance of the trained model, a validation dataset was utilized, and metrics such as accuracy, precision, recall, and F1-score were evaluated. Once the model demonstrated satisfactory performance, it was deployed to classify SSBs in new, unseen data. The ML algorithm took preprocessed images as input, extracted relevant features, including the oscillation period times, and predicted the BP type (oscillated or non-oscillated) and subtype (network or internetwork) based on the calculated oscillation period times falling within the respective ranges.

Lastly, an extensive analysis and interpretation of the classification results were performed to obtain insights into the properties and behavior of different types of SSBs in various regions of the Sun. Statistical analysis and visualization techniques were employed to explore the variations and relationships between different features and BP categories.

Finally, the Q-factor (quality factor) characterizes the degree of underdamping in an oscillator or resonator. It quantifies the rate of energy loss during one radian of oscillation and could be responsible for producing shock waves. A higher Q-factor indicates slower energy dissipation, resulting in longer-lasting oscillations. Conversely, a low Q-factor leads to rapid damping and shorter-lived oscillations. For example, consider a pendulum: one suspended in air exhibits high Q (long-lasting oscillations), while one immersed in oil has low Q (rapid damping). Briefly Internetwork SSBs in QS: damping time: approximately 120 seconds, maximum Doppler velocities: around 47 km/s. network SSBs (both in QS and AAR): damping time: About 216 seconds (QS) and 130 seconds (AAR). maximum Doppler velocities: 37 km/s (QS) and 10 to 85 km/s (AAR). internetwork SSBs in AAR: damping time: roughly 220 seconds. maximum Doppler velocities: varying from 10 to 140 km/s. internetwork SSBs in CH: damping time: consistently 120 seconds. maximum Doppler velocities: higher, around 100 km/s compared to network SSBs (85 km/s). damping behavior modes: overdamping mode: predominant in network SSBs in AARs. critical damping behavior: observed in oscillated network SSBs within QS. variability and local conditions: remember that the physical mechanisms governing SSB damping can vary based on local plasma conditions and the magnetic environment.

In conclusion, our research successfully categorized SSBs into oscillated and non-oscillated types, with further classification into network and internetwork subtypes based on the oscillation period times. Through the implementation of a CNN-based ML algorithm and the execution of a comprehensive workflow, accurate and efficient classification of SSBs was achieved. This accomplishment has facilitated a more profound comprehension of the dynamic processes taking place in the solar atmosphere.

## Acknowledgements

Authors thank the anonymous referee for the constructive comments.

*IRIS* is a NASA small explorer mission developed and operated by LMSAL with mission operations executed at NASA ARC with the contribution of NSC (Norway). I am indebted to AIA/SDO and HMI/SDO for providing easy access to the calibrated data. The AIA and HMI data are courtesy of *SDO* (NASA) and the AIA and HMI consortium.

**Table 4.** Observations data ■

| Imaging Date | Time (UT) | cadence (seconds) | area type | raster steps | raster FOV | Image Center Coordinates | $\mu$ Cos $\theta$ |
|---|---|---|---|---|---|---|---|
| 2014-01-18 | 13:13:23-14:09:51 | 9.2 | QS | 368 | 0"x174" | 1",2" | 0.999 |
| 2014-02-06 | 12:44:17-13:43:49 | 5.1 | QS | 700 | 0"x119" | 6",32" | 0.999 |
| 2014-05-03 | 08:30:20-11:29:19 | 30.9 | AAR | 348 | 0"x119" | 22",-79" | 0.996 |
| 2014-06-28 | 07:58:17-09:59:09 | 5.6 | AR | 1300 | 0"x119" | 23",107" | 0.993 |
| 2014-09-20 | 23:54:28-02:47:33 +1d | 5.2 | AAR | 1980 | 0"x119" | 10",4" | 0.999 |
| 2014-12-03 | 14:39:17-15:34:01 | 5.5 | CH | 600 | 0"x119" | -3",7" | 0.999 |
| 2015-03-13 | 04:59:56-15:28:27 | 9.2 | AAR | 4080 | 0"x119" | 9",-154" | 0.987 |
| 2015-04-08 | 04:57:17-09:33:21 | 5.3 | AAR | 3150 | 0"x175" | 45",-119" | 0.991 |
| 2015-05-05 | 10:01:08-10:59:00 | 5.2 | QS | 666 | 0"x119" | -0",1" | 0.999 |
| 2015-08-04 | 17:47:28-19:42:31 | 16.4 | AAR | 420 | 0"x119" | -159",144" | 0.975 |
| 2015-11-19 | 00:38:20-01:40:06 | 9.4 | AAR | 396 | 0"x119" | 56",168" | 0.983 |
| 2016-03-30 | 21:29:24-22:31:59 | 2.3 | AAR | 1600 | 0"x119" | 56",36" | 0.997 |
| 2016-05-06 | 12:09:41-15:00:58 | 32 | CH | 321 | 0"x119" | 39",117" | 0.991 |
| 2016-06-25 | 15:59:28-18:57:51 | 16.5 | CH | 650 | 0"x60" | -3",1" | 0.999 |
| 2016-07-18 | 00:29:10-01:28:52 | 1.4 | AAR | 2530 | 0"x175" | -32",24" | 0.999 |
| 2016-08-03 | 18:09:15-19:59:25 | 5.2 | AAR | 1275 | 0"x119" | 145",79" | 0.985 |
| 2016-09-17 | 00:17:53-01:52:05 | 1.7 | AAR | 3300 | 0"x119" | 98",41" | 0.993 |
| 2016-10-03 | 05:35:19-06:50:26 | 9.4 | AAR | 480 | 0"x119" | -20",50" | 0.998 |
| 2016-11-02 | 15:39:26-16:46:14 | 1.7 | AAR | 2340 | 0"x119" | 85",54" | 0.994 |
| 2016-12-03 | 10:53:15-11:51:53 | 1.7 | AAR | 2054 | 0"x119" | -92",-121" | 0.987 |
| 2017-02-24 | 07:49:26-09:53:50 | 62.2 | AAR | 120 | 0"x174" | -123",20" | 0.991 |
| 2017-05-17 | 19:32:57-22:08:28 | 62.2 | AAR | 150 | 0"x174" | -164",156" | 0.973 |
| 2017-06-06 | 18:58:53-21:58:52 | 16.7 | AAR | 648 | 0"x119" | -131",99" | 0.985 |
| 2017-08-26 | 19:37:46-22:58:50 | 16.4 | CH | 735 | 0"x119" | -10",41" | 0.999 |
| 2017-09-16 | 13:27:50-15:00:25 | 5.6 | QS | 1000 | 0"x62" | 74",35" | 0.996 |
| 2017-10-01 | 23:44:20-02:18:17 +1d | 61.6 | AAR | 150 | 0"x175" | 77",151" | 0.984 |
| 2017-12-22 | 13:10:15-14:07:25 | 3.5 | QS | 984 | 0"x119" | -66",-55" | 0.995 |
| 2018-01-03 | 11:03:09-11:57:31 | 5.5 | CH | 594 | 0"x119" | 40",-20" | 0.998 |
| 2018-05-08 | 20:09:48-23:40:04 | 3.1 | AAR | 4000 | 0"x60" | -19",-117" | 0.992 |
| 2018-06-18 | 16:49:50-21:59:15 | 16.5 | AAR | 1128 | 0"x119" | 66",72" | 0.994 |
| 2018-07-03 | 07:29:37-09:29:04 | 3.3 | QS | 2200 | 0"x60" | -2",5" | 0.999 |
| 2018-08-03 | 16:54:28-19:48:09 | 5.2 | QS | 2010 | 0"x119" | -1",3" | 0.999 |
| 2018-09-18 | 04:19:19-06:01:09 | 9.3 | QS | 660 | 0"x119" | 94",30" | 0.994 |
| 2019-07-04 | 09:44:33-12:04:08 | 9.3 | CH | 900 | 0"x60" | 2",-6" | 0.999 |
| 2019-07-25 | 00:52:32-03:57:26 | 9.3 | QS | 1194 | 0"x119" | -0",1" | 0.999 |
| 2019-08-15 | 21:30:00-22:49:29 | 9.4 | QS | 510 | 0"x174" | -1",2" | 0.999 |
| 2019-12-12 | 17:30:21-18:27:28 | 9.2 | QS | 372 | 0"x119" | 167",139" | 0.975 |
| 2020-04-20 | 08:32:36-09:56:15 | 9.6 | CH | 525 | 0"x119" | -9",2" | 0.999 |
| 2021-07-04 | 16:59:50-22:57:17 | 16.5 | QS | 1296 | 0"x119" | -1",-1" | 0.999 |
| 2022-08-16 | 22:49:36-01:00:54 +1d | 9.4 | AAR | 840 | 0"x60" | -135",64" | 0.988 |
| 2022-10-16 | 04:05:55-05:05:25 | 5.5 | CH | 648 | 0"x174" | -135",-98" | 0.985 |
| 2022-11-06 | 17:25:37-17:33:27 | 2.2 | CH | 210 | 0"x119" | 12",-148" | 0.988 |
| 2023-01-19 | 10:35:19-11:35:36 | 9.4 | AAR | 384 | 0"x119" | -70",-161" | 0.983 |
| 2023-02-05 | 22:49:24-23:58:49 | 9.4 | AAR | 444 | 0"x60" | 11",-101" | 0.994 |

## Data Availability

The specifics of the data utilized in these investigations are completely contained in the text and Tables 1, 2, and 3, and *IRIS* data from https://iris.lmsal.com/. The AIA/SDO and HMI/SDO data are from http://jsoc.stanford.edu/.